\documentclass[aps,prl,reprint,twocolumn]{revtex4-1}

\usepackage{xcolor}
\usepackage{graphicx}
\usepackage[utf8]{inputenc}
\usepackage{amsmath}
\usepackage{amssymb}

\newcommand{\ket}[1]{\big|#1\big>}

\newcommand{\bk}{\mathbf{k}}

\newcommand{\bq}{\mathbf{q}}

\def\dz2{d$_{\text{z}^2}$}

\def\dx2y2{d$_{\text{x}^2\text{y}^2}$}
\def\G0W0{G$_0$W$_0$}
\def\scGW0{scGW$_0$}

\def\mos{MoS$_2$}
\def\mose{MoSe$_2$}
\def\ws{WS$_2$}
\def\wse{WSe$_2$}
\def\sio{SiO$_2$}

\begin{document}

\title{Excitation-induced transition to indirect band gaps in atomically thin transition metal dichalcogenide semiconductors}

\author{D. Erben$^{1}$}
\email{derben@itp.uni-bremen.de}
\author{A. Steinhoff,$^{1}$ G. Sch\"onhoff,$^{1,2}$ T.O. Wehling,$^{1,2,3}$ C. Gies$^{1}$}
\author{F. Jahnke$^{1,3}$}

\address{$^1$Institut f\"ur Theoretische Physik, Universit\"at Bremen, P.O. Box 330 440, 28334 Bremen, Germany}
\address{$^2$Bremen Center for Computational Materials Science, Universit\"at Bremen, 28334 Bremen, Germany}
\address{$^3$MAPEX Center for Materials and Processes, Universit\"at Bremen, 28359 Bremen, Germany}

%
%
%
%

%

\begin{abstract}

Monolayers of transition metal dichalcogenides (TMDCs) exhibit an exceptionally strong Coulomb interaction between charge carriers due to the two-dimensional carrier confinement in connection with weak dielectric screening. High densities of excited charge carriers in the various band-structure valleys cause strong many-body renormalizations that influence both the electronic properties and the optical response of the material. We investigate electronic and optical properties of the typical monolayer TMDCs \mos{}, \mose{}, \ws{} and \wse{} in the presence of excited carriers by solving semiconductor Bloch equations on the full Brillouin zone. With increasing carrier density, we systematically find a reduction of the exciton binding energies due to Coulomb screening and Pauli blocking. Together with excitation-induced band-gap shrinkage this leads to redshifts of excitonic resonances up to the dissociation of excitons. As a central result, we predict for all investigated monolayer TMDCs that the $\Sigma$-valley shifts stronger than the K-valley. Two of the materials undergo a transition from direct to indirect band gaps under carrier excitation similar to well-known strain-induced effects. Our findings have strong implications for the filling of conduction-band valleys with excited carriers and are relevant to transport and optical applications as well as the emergence of phonon-driven superconductivity.

\end{abstract}
\maketitle
\section{Introduction}
Monolayers of transition metal dichalcogenide (TMDC) semiconductors can be used as a novel active material in optoelectronic devices such as ultrasensitive photodetectors \cite{lopez-sanchez_ultrasensitive_2013}, light-emitting diodes \cite{pospischil_solar-energy_2014, baugher_optoelectronic_2014, ross_electrically_2014, withers_light-emitting_2015}, solar cells \cite{pospischil_solar-energy_2014, baugher_optoelectronic_2014}, and lasers \cite{wu_monolayer_2015, ye_monolayer_2015, salehzadeh_optically_2015, li_room-temperature_2017}. Of fundamental interest is also the possibility to combine individual two-dimensional materials in functional van der Waals-heterostructures \cite{geim_van_2013}, and the selective optical addressability of band-structure valleys as a new degree of freedom. \cite{xu_spin_2014}
Fascinating prospects arise from the possibility to engineer electronic and optical properties by manipulation of the Coulomb interaction in atomically thin materials \cite{latini_excitons_2015, steinke_noninvasive_2017, rosner_two-dimensional_2016, raja_coulomb_2017, steinhoff_exciton_2017}. Modifications of the dielectric environment as well as the application of strain can change the band structure by hundreds of meV.
\par
An equally important, but less recognized source of screening is provided by charge carriers that are created by doping or excitation. Carrier doping in TMDCs is often described in terms of K and K' valleys, in the context of Fermi polarons \cite{efimkin_many-body_2017}, the valley Zeeman effect \cite{wang_strongly_2017}, optical properties \cite{chaves_excitonic_2017} or band-gap renormalization \cite{gao_renormalization_2017}. It is widely recognized that in tungsten-based compounds the conduction band corresponding to dark interband transitions is split off and drains carriers from the bright transition \cite{wang_spin-orbit_2015,echeverry_splitting_2016}. A frequently overlooked property of the $\Sigma$-valley, halfway between K and $\Gamma$, is that it may energetically shift \emph{below} the K-valley under carrier doping\cite{ge_phonon-mediated_2013} thus competing with the K-valley for excited carriers. The carrier density, at which this effect becomes relevant, is determined by the energetic distance of the valleys in the ground-state band structure of the material. The shift of the $\Sigma$-valley relatively to the K-valley facilitates a Lifshitz transition that enables phonon-driven superconductivity of monolayer TMDCs\cite{schonhoff_interplay_2016}. Therefore, a quantitative understanding of the valley shifts provides access to carrier-doping densities at which superconductivity is expected. 
The ``directness'' of the gap is also subject to strong variations depending on the lattice constant and strain \cite{steinhoff_efficient_2015} as well as the TMDC material. Therefore, under carrier excitation the $\Sigma$-valley may have a strong influence on the Fermi level, which is a central quantity especially at low temperatures, and therefore on transport properties and optoelectronic properties like gain \cite{chernikov_population_2015}.
Optical properties are not only affected by the single-particle electronic states, but also by two-particle interaction processes that yield a strong response due to excitons. Exciton binding energies as large as $0.6$ eV have been predicted and experimentally observed in MX$_{2}$ monolayer materials \cite{berkelbach_theory_2013, steinhoff_influence_2014, ugeda_giant_2014, mayers_binding_2015}. Phase-space filling and screening can drastically change the binding energy in a way that is different to band-structure renormalizations. \cite{steinhoff_influence_2014}
\par
In the present work, we provide systematic insight into the single- and two-particle properties of the four most commonly investigated TMDC semiconductors \mos{}, \mose{}, \ws{}, and \wse{} in the presence of excited carriers. We discuss conditions for a direct-to-indirect band-gap transition and provide signatures that can be used to identify this transition in experiments both relating to single-particle properties, such as photoemission spectroscopy, and optical properties via photoluminescence. 
The direct-to-indirect transition is mostly driven by electron-hole exchange interaction among excited carriers. Electron-hole exchange is very strong in TMDC semiconductors \cite{qiu_nonanalyticity_2015}, yet up to now it has only been discussed in the context of exciton fine structure. \cite{qiu_nonanalyticity_2015,plechinger_trion_2016} Our results provide new insight into the validity of assuming a direct band gap for the monolayer TMDCs for given excitation scenarios and into the intrinsic differences between the four materials. We compare relative valley shifts obtained from a state-of-the-art many-body theory using a frequency-dependent GW self-energy, a self-energy in static approximation, and semi-local exchange-correlation potentials from DFT. Thereby we confirm that the direct-to-indirect transition is well-described already on the level of a statically screened self-energy.
\section{Modelling the excited-state properties of monolayer TMDCs}
Our theoretical description of TMDC monolayers uses the formalism of semiconductor Bloch equations (SBE) in combination with ab-initio calculations for the ground-state properties. These equations provide access to the optical properties of semiconductors based on material-realistic band structures and interaction matrix elements and are able to describe effects of excited charge carriers on the materials' response. Scattering processes due to carrier-carrier Coulomb or carrier-phonon interaction can systematically be included to account for dephasing of optical transitions.
In this work we aim for an investigation of the charge carrier distribution, band-structure renormalizations, and optical properties in response to a weak optical probe field. We do not consider dynamics of charge-carrier populations, but focus on the time window following the excitation and relaxation of electrons and holes, e.g., by an ultra-short laser pulse preceeding the probe pulse. For a sufficiently long delay between pump and probe pulse, the latter will record a response to a thermalized charge-carrier distribution with a given density and temperature before recombination sets in.
\par
The SBE including interaction-induced dephasing due to carrier-carrier-interaction, described in GW-approximation, are derived in the Appendix. There we also outline how to obtain the static limit, or \textit{screened-exchange Coulomb-hole} approximation (SXCH), of the theory, which considerably simplifies the numerical evaluation.
Our theoretical description of the optical properties of TMDC semiconductors is based on equations of motion for the electron-hole interband transition amplitudes $\psi^{he}_{\bk}(t)$, which are driven by the weak optical probe field $\mathbf{E}(t)$
%
\begin{equation}
 \begin{split}
 i\hbar\frac{\mathrm{d}}{\mathrm{d}t}\psi_{\bk}^{he}(t) &= 
 \left(\tilde{\varepsilon}_{\bk}^{h}+\tilde{\varepsilon}_{\bk}^{e}-i\gamma\right)\psi_{\bk}^{he}(t) \\
 -\Big(\mathbf{d}_{\bk}^{eh}\cdot \mathbf{E}(t)&+\frac{1}{\mathcal{A}}\sum_{\bk'}W_{\bk\bk'\bk\bk'}^{ehhe}\psi_{\bk'}^{he}(t)\Big) \left(1-f_{\bk}^{e}-f_{\bk}^{h}\right).
 \end{split}
\label{eq:SBE_main}
\end{equation}
Here $\mathbf{d}^{eh}_{\bk}$ is the dipole coupling matrix element, $W_{\bk\bk'\bk\bk'}^{ehhe}$ is the screened Coulomb interaction matrix element, $f^{e,h}_{\bk}$ are the population functions of excited electrons and holes, and $\gamma$ is a dephasing constant. The renormalized energies $\tilde{\varepsilon}_{\bk}^{\lambda}=\varepsilon_{\bk}^{0,\lambda}+\Sigma_{\bk}^{\textrm{H},\lambda}+\Sigma_{\bk}^{\textrm{U},\lambda}+\Sigma_{\bk}^{\textrm{SX},\lambda}+\Sigma_{\bk}^{\textrm{CH},\lambda}$ are given in the Appendix. 
Band structures $\varepsilon_{\bk}^{0,\lambda}$ of the optically relevant lowest conduction and highest valence bands are obtained from a \G0W0{}-calculation as described in Ref.~\citenum{steinhoff_influence_2014}. The valence- and conduction-band splitting caused by spin-orbit interaction is considered along the lines of Ref.~\citenum{liu_three-band_2013} and \citenum{steinhoff_influence_2014}, including first- and second-order effects. To take into account dielectric screening by the environment, we use the \textit{Wannier function continuum electrostatic} (WFCE) approach described in Ref.~\citenum{rosner_wannier_2015} that combines a continuum-electrostatic model for the screening with a localized description of Coulomb interaction provided in Ref.~\citenum{steinhoff_exciton_2017}.
The band structure is renormalized due to many-body Coulomb interaction among excited carriers. Screening of the Coulomb interaction by excited electrons and holes is treated in the long-wavelength limit, see the Appendix. 
In the same way as the single-particle properties, light-matter interaction is renormalized by many-body Coulomb effects. This is described in Eq.~(\ref{eq:SBE_main}) via the coupling of the interband transition amplitudes $\psi^{he}_{\bk}$ for different carrier momenta $\bk$ mediated by the screened Coulomb interaction. This coupling gives rise to the presence of excitonic resonances in the optical response. In this sense, single-particle and two-particle properties act together in the materials' optical response that is accessible via the macroscopic polarization $P(t)=\sum_{\bk,h,e}\left[\psi_{\bk}^{he}(t)(\mathbf{d}_{\bk}^{eh})^*+c.c.\right]$. From the Fourier transform of P(t) and the weak optical probe field E(t), the imaginary part of the linear optical susceptibility $\chi(\omega)=P(\omega)/E(\omega)$ provides access to the absorption spectrum.\\
In the next two sections we focus on the single- and two-particle results, respectively. Starting with single-particle properties, we take a closer look at band-structure renormalizations and single-particle populations in the full Brillouin zone. Proceeding to two-particle observables, we analyze exciton binding energies and optical absorption spectra.
\begin{figure*}[h!t]
\centering
\includegraphics[trim = 5.0cm 5.0cm 5.0cm 5.0cm, width=\textwidth]{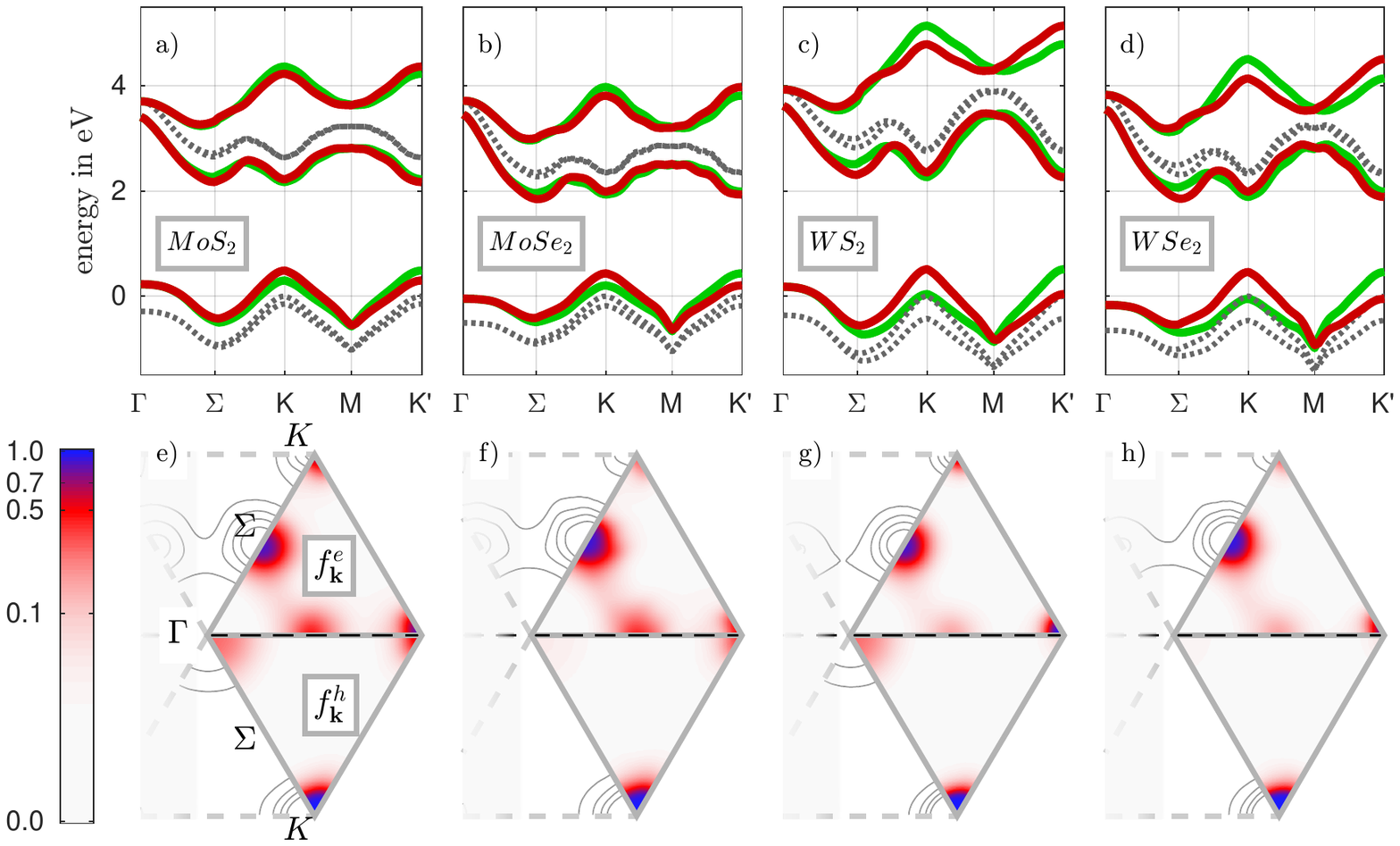}
\caption{Renormalized band structures of \mos{}, \mose{}, \ws{}, and \wse{} at $T=300$ K are presented in the top row (a) - (d). The zero-density band structure is shown as grey dashed line indicating huge changes in the presence of excited carriers. The bottom row (e) - (h) shows the electron (hole) population in the lowest conduction band (highest valence band) for the corresponding materials at a density of $3.2\times 10^{13}$ cm$^{-2}$. High occupations at both K and $\Sigma$ are clearly visible. Moreover, the difference in the occupation of K/K' and $\Sigma$/$\Sigma'$ between molybdenum and tungsten compounds as well as the sulfide/selenide compounds can be seen.} 
\label{fig:bs_fk}
\end{figure*}
\begin{figure}[h!t]
\centering
\includegraphics[width=.8\columnwidth]{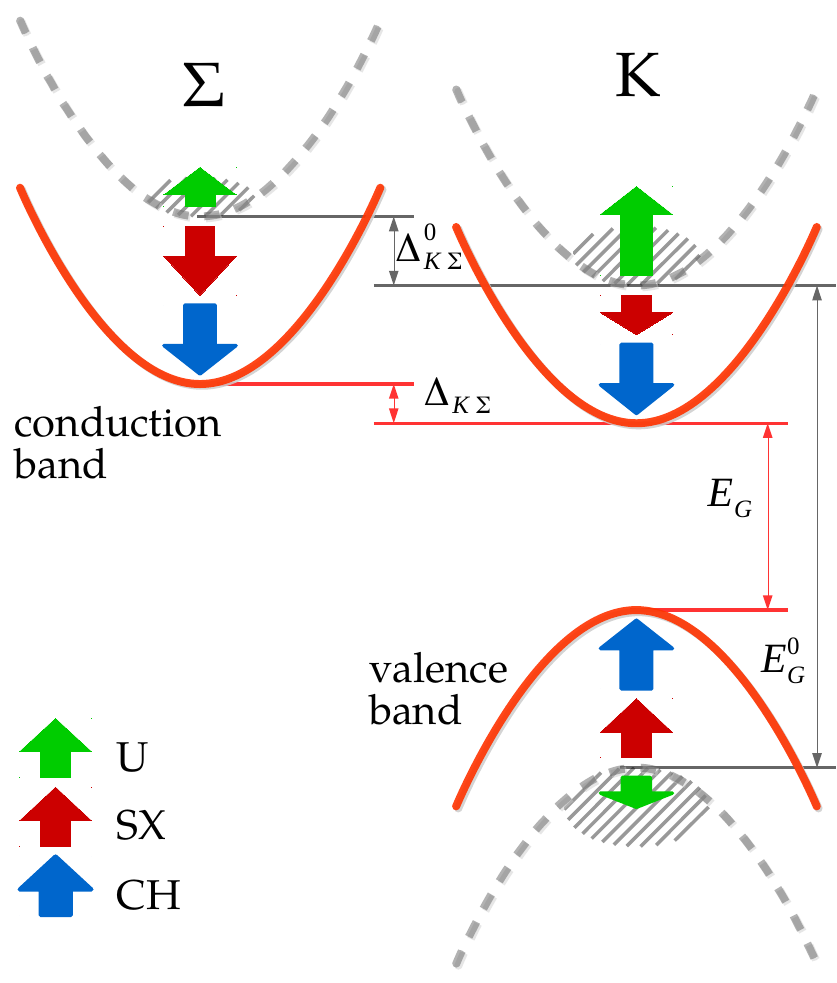}
\caption{Schematic illustrating the effect of excited carriers (shaded areas): Shown are the band-structure renormalization at K and $\Sigma$ for spin-up carriers by intraband screened-exchange (SX), Coulomb-hole (CH) and interband electron-hole exchange (U) shifts. The situation is equivalent for spin-down carriers at K'- and $\Sigma'$-valleys.}
\label{fig:scheme}
\end{figure}
\begin{figure}[h!t]
\centering
\includegraphics[width=.35\textwidth]{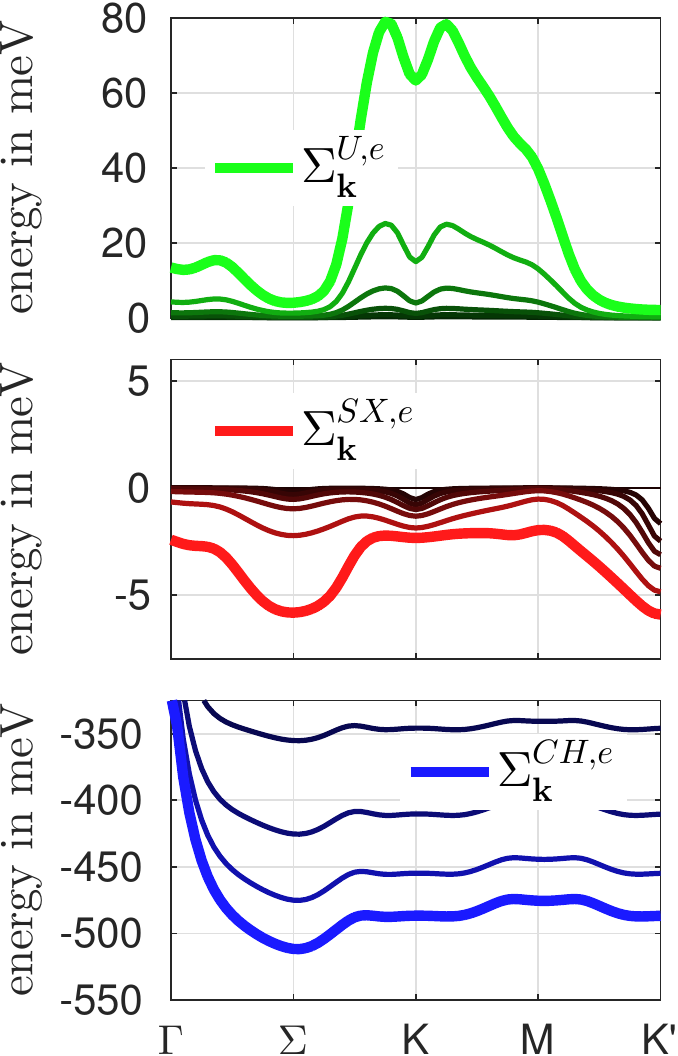}
\caption{Renormalization shifts of the spin-up conduction band for \ws{} along the $\Gamma$-K-K' path at $T=300$ K. Bare electron-hole exchange, screened-exchange and Coulomb-hole shifts are separately shown (from top to bottom) for increasing charge-carrier densites $(1.0\times10^{11}, 3.2\times10^{11},1.0\times10^{12},3.2\times10^{12},1.0\times10^{13},3.2\times10^{13}$ cm$^{-2} )$. The screened electron-electron exchange interaction (SX) shifts $\Sigma$ downwards more strongly than K, while the unscreened electron-hole exchange (U) pushes K upwards. Note the different energy scales.}
\label{fig:sx_ch}
\end{figure}
%
In the following, we analyze the band-structure renormalizations and populations in freestanding \mos{}, \mose{}, \ws{} and \wse{} monolayers as a function of the electron-hole pair densities. We further consider the case of either electron or hole doping.
As a general trend, renormalization effects lower the conduction bands and raise the valence bands, as described by the SXCH self-energy given in Eq. (\ref{eq:en_SXCH}). More subtle yet important details are revealed from the momentum dependence of these effects. First, one has to consider the differences between the band structures of the four materials in their ground state and the implications for quasi-thermal populations of carriers in the Brillouin zone. As shown in Fig.~\ref{fig:bs_fk}, \mose{} and \wse{} have an intrinsically indirect band gap at the K-point. This leads to a drain of charge carriers from K to $\Sigma$, see Fig.~\ref{fig:bs_fk}f and \ref{fig:bs_fk}h. In comparison, \mos{} and \ws{} show a direct band gap, although the indirect band gap at $\Sigma$ has only a slightly larger gap energy. This leads to almost equal populations in the conduction-band K and $\Sigma$-valleys. On the other hand, both tungsten-based TMDCs have a larger energy difference between $\
\Sigma/\Sigma'$ as well as between K/K', which results in small occupancies of $\Sigma'$ in \ws{} and \wse{}.
Additionally, one has to consider the different spin-orbit splitting for \mos{} and \ws{}. For \ws{} the conduction-band splitting is larger than for \mos{} and also of opposite sign, which means that the lowest possible transition in \ws{} is spin-forbidden. This results in intrinsic loss of electrons at K for bright transitions in tungsten-based compounds.
As no $\Sigma$-valley exists in the valence band, the holes gather solely at K and K'. For \mos{} and \ws{} there are small hole occupancies in the $\Gamma$-valleys as well, as the $\Gamma$-point is renormalized more strongly than the K-point similar to tensile-strain effects \cite{conley_bandgap_2013,steinhoff_efficient_2015}. However, this fact is less relevant than for the $\Sigma$-point, as the initial separation between K and $\Gamma$ is larger.
High occupancies at K and $\Sigma$ have been observed in several recent experiments using angular-resolved photoemission spectroscopy (ARPES) \cite{grubivsic_vcabo_observation_2015, bertoni_generation_2016, ulstrup_spin_2017}. Band-gap and binding-energy shrinkage on the order of several $100$ meV observed in optical spectroscopy have also been reported\cite{cunningham_photoinduced_2017}.
\par
Eq.~(\ref{eq:en_SXCH}) allows us to reveal the mechanisms that are responsible for the stronger shift at $\Sigma$ compared to K and the resulting direct-to-indirect band-gap transition for \mos{} and \ws{}. Two mechanisms provide momentum-dependent energy renormalizations affecting screened-exchange and Coulomb-hole contributions (SX and CH, intraband interaction) as well as the interband electron-hole-exchange separately as illustrated in Fig.~\ref{fig:scheme}. Hartree renormalizations turn out to be small compared to all other contributions and are thus not discussed explicitely. The first mechanism arises from the valley dependence of the involved Coulomb matrix elements 
\begin{equation}
V^{\lambda \lambda' \lambda \lambda'}_{\bk_1 \bk_2 \bk_3 \bk_4} = \sum_{\alpha, \beta} \Big( c_{\alpha, \bk_1}^{\lambda} \Big)^{*} \Big( c_{\beta, \bk_2}^{\lambda'} \Big)^{*} c_{\beta, \bk_3}^{\lambda} c_{\alpha, \bk_4}^{\lambda'} V^{\alpha\beta\beta\alpha}_{|\bk_1-\bk_4|}\,,
\label{eq:v_abcd}
\end{equation}
which are obtained directly from \G0W0{}-calculations in a localized basis of Wannier functions $\ket{\alpha}$ with dominant transition-metal d-orbital character \cite{liu_three-band_2013, steinhoff_influence_2014}.
\begin{figure*}[h!t]
\centering
\includegraphics[width=\textwidth]{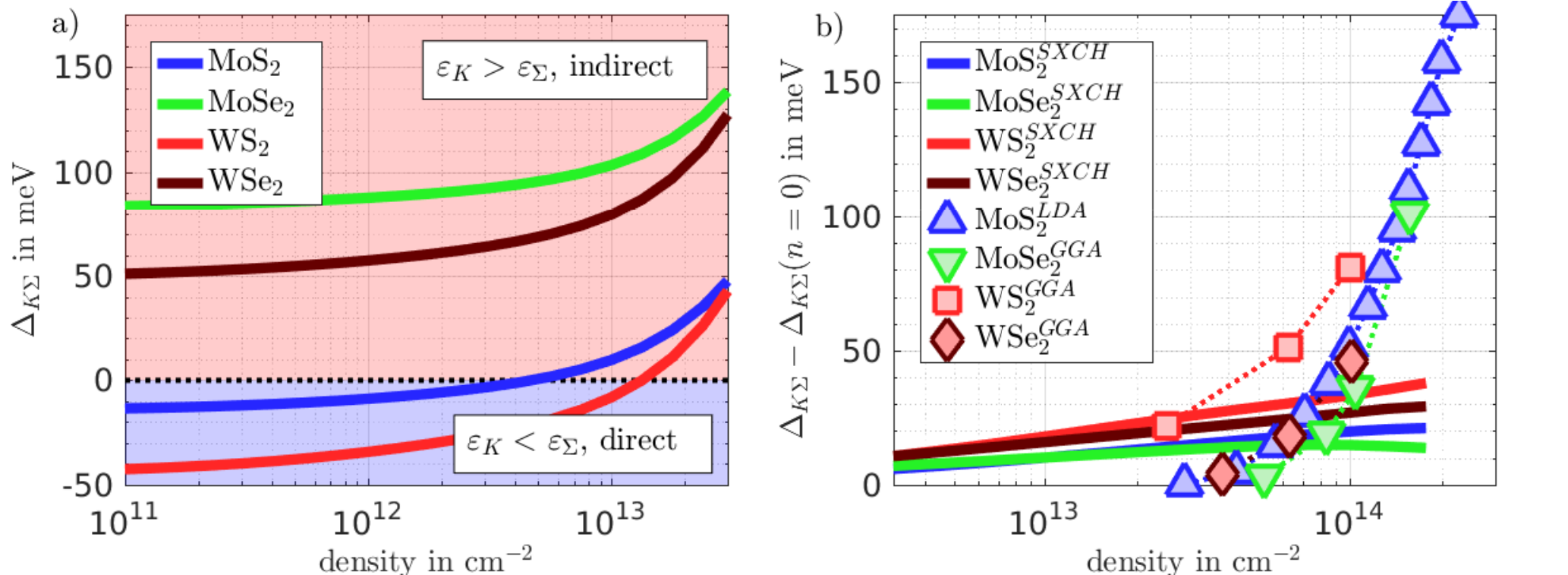}
\caption{\textbf{a)} Energy difference between the K- and $\Sigma$-valley at $T=300$ K for photodoping with electrons and holes obtained in SXCH approximation plus electron-hole exchange. A clear trend of $\Sigma$ shifting energetically below K (increasing $\Delta_{K \Sigma}$) is visible for all four materials. \textbf{b)} Energy difference between the K- and $\Sigma$-valley at $T=300$ K relative to the intrinsic energy difference at zero carrier density for doping with electrons. We compare results obtained in SXCH approximation plus electron-hole exchange (solid lines) to data obtained from density functional theory (DFT, dashed lines with symbols) using different approximations (LDA/GGA) for the exchange-correlation functionals.}
\label{fig:deltaKS}
\end{figure*}
\begin{figure}[h!t]
\centering
\includegraphics[width=\columnwidth]{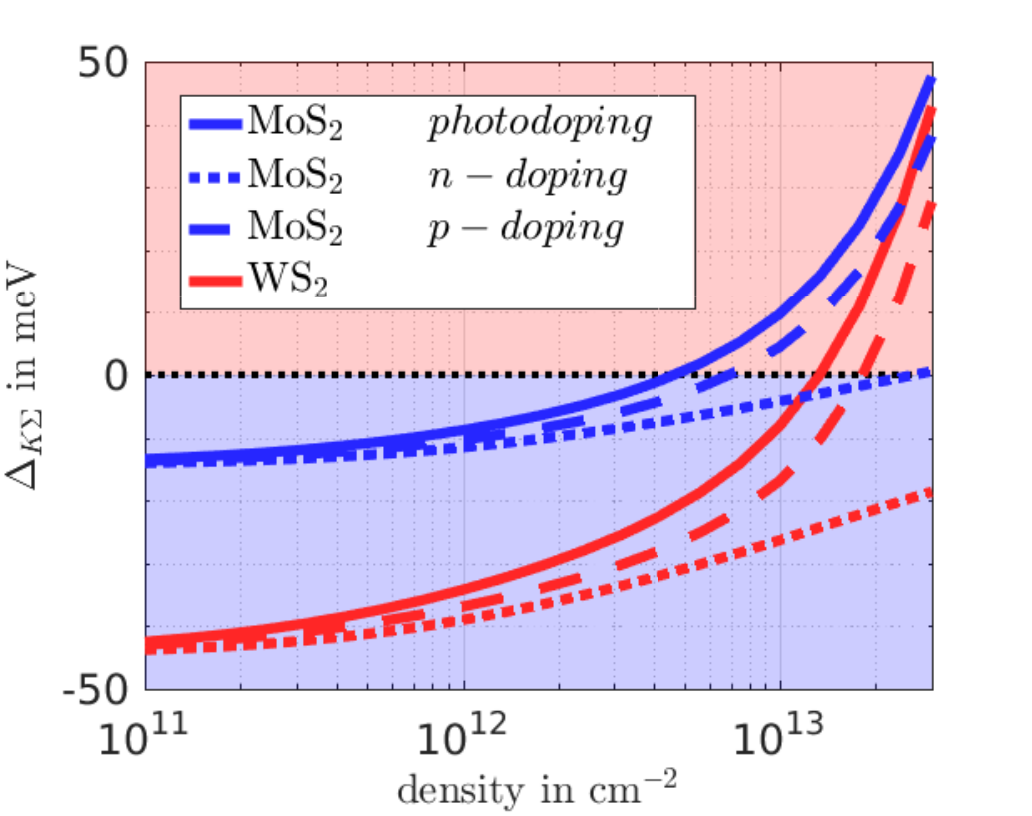}
\caption{Energy difference between the K- and $\Sigma$-valley at $T=300$ K comparing photodoping with electrons and holes (solid lines), electron doping (n-doping, dotted lines) and hole doping (p-doping, dashed lines). The results are obtained in SXCH approximation plus electron-hole exchange.}
\label{fig:deltaKS_02}
\end{figure}
Differing orbital characters of the involved Bloch states in the K- and $\Sigma$-valleys ($d_{z^2}$ at K, $d_{x^2-y^2}$ and $d_{xy}$ at $\Sigma$) cause a valley sensitivity of the resulting energy renormalizations, which are stronger in the $\Sigma$-valley.
The second mechanism is driven by the valley dependence of excited-carrier populations. The screened intra-band exchange energy $\Sigma_{\bk}^{\textrm{SX},\lambda}$ as well as the unscreened electron-hole exchange energy $\Sigma_{\bk}^{\textrm{U},\lambda}$ given by
are proportional to the carrier populations shown in Fig.~\ref{fig:bs_fk}. The absence of hole population outside the K-valley implies that a weighting of Coulomb matrix elements by hole populations occurs only at K. This means that the redshift due to electron-electron exchange interaction at $\Sigma$ is barely compensated by any blueshift due to electron-hole exchange interaction, while at K electron-hole exchange yields a strong blueshift. The latter is favored by the fact that electron-hole exchange is practically insensitive to screening, see the discussion in the Appendix. As a result for the conduction band, additionally to the more or less rigid Coulomb-hole shift, $\Sigma$ and K' are shifted downwards and K is shifted upwards. Numerical results of this behaviour are shown in Fig.~\ref{fig:sx_ch}. On the other hand, valence-band renormalization occurs mostly in the vicinity of the K-valley. Here, the upper valence band (lower hole band) is renormalized more strongly than the lower valence band due to larger hole populations, which leads to an enhancement of the effective spin-orbit splitting at the K point. The enhancement amounts to approximately $40$ meV for all four materials at a hole density of $3.2\times 10^{13}$ cm$^{-2}$.
As Fig.~\ref{fig:sx_ch} shows, our calculations predict renormalizations on the order of $500$ meV for elevated densities. There is an increasing shift at $\Sigma$ relative to K in the conduction band due to the reasons explained above, the energy difference being labeled $\Delta_{K \Sigma}$. As summarized in Fig.~\ref{fig:deltaKS}a, all materials show the same tendency to become (more) indirect semiconductors with increasing excitation density. This strongly influences optical properties like photoluminescence yield and gain, which are both sensitive to the combined populations of electrons and holes at the direct band gap.\cite{steinhoff_efficient_2015, chernikov_population_2015}
\par
The relative valley shifts are not only relevant for optical properties but also determine the onset of superconductivity in electron-doped TMDCs. It is believed\cite{ge_phonon-mediated_2013} that phonon-driven superconductivity sets in as soon as both, the K- and $\Sigma$-valley become populated with electrons. In this context the shifts of conduction-band valleys in n-doped TMDCs have so far been analyzed using DFT \cite{ge_phonon-mediated_2013, schonhoff_interplay_2016}. A comparison of the results we obtained in SXCH approximation plus electron-hole exchange to DFT calculations in LDA and GGA, respectively, for the case of n-doping is shown in Fig.~\ref{fig:deltaKS}b. We observe that the SXCH approximation predicts stronger relative valley shifts at low densities, while the shifts obtained from DFT calculations become stronger at densities above $5\times 10^{13}$ cm$^{-2}$. A central reason for this discrepancy is that screening due to excited carriers is treated differently in DFT and the diagrammatic SXCH approximation. The strong impact of electron-hole exchange on the energy difference $\Delta_{K \Sigma}$ is illustrated in Fig.~\ref{fig:deltaKS_02}, where we directly compare the effects of photodoping, n-doping and p-doping. 
In n-doped materials, the exchange interaction always leads to redshifts of the band structure as there is no additional positive shift at K due to the missing holes in the valence band. Hence the overall direct-to-indirect effect is weakened compared to photo-doped materials.
On the other hand, the effect is more pronounced for p-doped materials, as there is a large positive shift at K in the conduction band due to the interband electron-hole exchange but no screened-exchange shift due to electrons. The results suggest that an initially n-doped monolayer TMDC might by driven beyond the Lifshitz transition by slight photodoping due to the strong electron-hole exchange. We thus speculate that photodoping might be a means to optically induce superconductivity in atomically thin TMDCs. 
%
\begin{figure}[h!t]
\centering
\includegraphics[width=1.0\columnwidth]{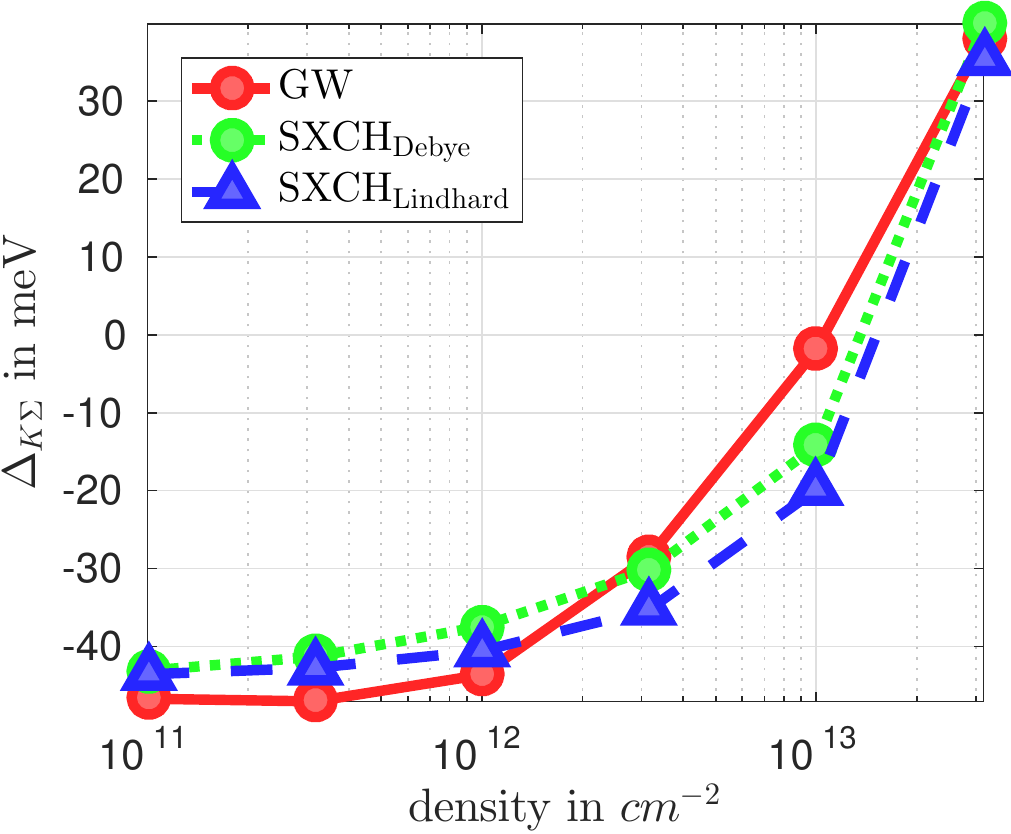}
\caption{Difference of K- and $\Sigma$-valley conduction-band energies ($\Delta_{K \Sigma}$, see Fig.~\ref{fig:scheme}) at $T=300$ K in monolayer WS$_2$ on SiO$_2$ substrate under carrier excitation obtained on different levels of approximation. The solid red line shows results from GW-calculations including dynamical screening, in comparison to the results from static SXCH-theory including screening in Lindhard (dashed blue line) and in Debye (long-wavelength, dotted green line) approximation, see the Appendix.}
\label{fig:GW_vs_SXCH}
\end{figure}
\par
All results discussed so far are obtained treating carrier-carrier interaction in static (SXCH) approximation. Calculating band-structure renormalizations on the level of a frequency-dependent GW approximation (e.g. including dynamical screening) according to Eqs.~(\ref{eq:quasip}) and (\ref{eq:MW}) increases the numerical effort significantly. A comparison of results from static and frequency-dependent calculations reveals that the direct-to-indirect transition is also observed in the more elaborate theory, see  Fig.~\ref{fig:GW_vs_SXCH}. Both, static and full GW-approximation of the Coulomb interaction lead to stronger renormalizations at $\Sigma$ than at K, as the main source of this effect is the bare electron-hole exchange. The overall quantitative agreement is very good, which confirms the validity of the static approximation.
%
%
%
%
%
%
\section{Two-particle properties}
\begin{figure*}[h!t]
\centering
\includegraphics[width=\textwidth]{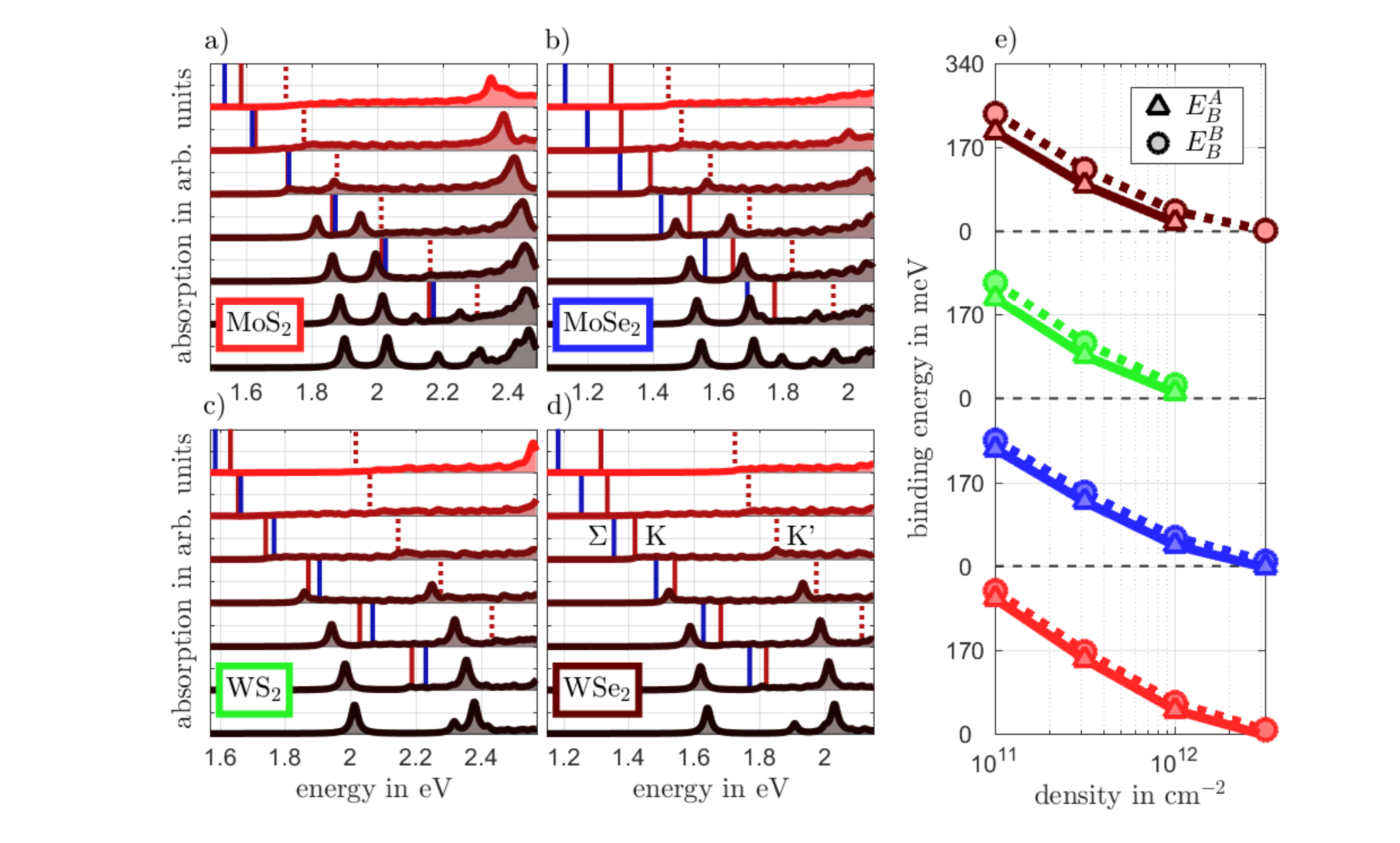}
\caption{ \textbf{a) - d)}: Absorption spectra for various TMDCs at $T=300$ K and increasing electron-hole pair density from bottom to top $( \textrm{ground state}, 1.0\times10^{11}, 3.2\times10^{11}, 1.0\times10^{12}, 3.2\times10^{12} ,1.0\times10^{13}, 3.2\times10^{13}$ cm$^{-2} )$. Due to the over-estimation of the \G0W0{}-band-gap the spectra are shifted such that the exciton peak positions at zero density agree with experiment. The vertical lines indicate the quasiparticle band gaps at K ($\tilde{\varepsilon}_K^h \rightarrow \tilde{\varepsilon}_K^e$, direct gap, solid red), $\Sigma$ ($\tilde{\varepsilon}_K^h \rightarrow \tilde{\varepsilon}_\Sigma^e$, indirect gap, solid blue) and K' ($\tilde{\varepsilon}_{K'}^h \rightarrow \tilde{\varepsilon}_{K'}^e$, direct gap, dashed red) for spin-up carriers. For \mos{} and \ws{} the indirect band gap at $\Sigma$ becomes comparable to or even smaller than the direct band gap at K, the material thus becoming indirect due to band-structure renormalizations. \textbf{e)}: Binding energies of A and B excitons in dependence of the charge carrier density. The vanishing of binding energies ($E^X_B < 0$) indicates the exciton dissociation (Mott transition).}
\label{fig:spectra_eb}
\end{figure*}

The full solution of the SBE combines the single-particle renormalizations with the  attractive electron-hole interaction to yield the absorption spectra shown in Fig.~\ref{fig:spectra_eb}a to \ref{fig:spectra_eb}d. The A and B exciton lines separated by spin-orbit splitting are the dominating low-energy peaks. The excitons in general show a redshift with increasing carrier density as the result of two competing effects: on the one hand the quasi-particle band gap shrinks due to the above-discussed renormalizations (marked by vertical lines in the plots). On the other hand the exciton binding energy decreases due to screening of the Coulomb interaction as well as Pauli blocking in the presence of excited carriers as shown in Fig.~\ref{fig:spectra_eb}e. The exciton binding energies of freestanding monolayers on various substrates are presented in Tab.~(\ref{tbl:results}). We find them to be in the range of $0.5$ to $0.6$ eV in agreement with literature \cite{berkelbach_theory_2013, steinhoff_influence_2014, ugeda_giant_2014, mayers_binding_2015}. The Mott transition for the four materials can be determined from the zero-crossing of the density-dependent exciton binding energy. Our calculations yield densities around $3\times10^{12}$ cm$^{-2}$ for MoX$_{2}$ and $1\times10^{12}$ cm$^{-2}$ for WX$_{2}$. We note that recently slightly higher densities have been predicted using a many-body theory including frequency-dependent screening \cite{steinhoff_exciton_2017}, which indicates that the band-gap shrinkage obtained from a static calculation should be seen as an upper bound.

\par
For TMDC monolayers on a substrate, the renormalizations follow the same mechanisms as described above. Whereas for freestanding \ws{} we obtain a binding energy of $557$ meV, for \ws{} on \sio{} a drastical reduction to $268$ meV is found due to additional screening of Coulomb interaction by the substrate. This effect has been discussed in literature for TMDC layers in their ground state only \cite{lin_dielectric_2014,ugeda_giant_2014,latini_excitons_2015,trolle_model_2017,raja_coulomb_2017,cho_environmentally_2018}, while we provide results in the presence of excited carriers. The effect of the Coulomb-hole term becomes weaker since it involves the difference between Coulomb matrix elements screened and unscreened by excited carriers, both of which are reduced by environmental screening described by a dielectric function $\varepsilon_{\textrm{b}}^{-1}$: $\Sigma^{\textrm{CH}}\propto W-V=U\varepsilon_{\textrm{b}}^{-1}(\varepsilon_{\textrm{exc}}^{-1}-1)$. For \ws{} the band-gap shift reduces to a maximum value of about $570$ meV compared to $940$ meV for freestanding \ws{}. The difference in energy shift between K and $\Sigma$ is almost unaffected by environmental screening in the case of photo- or p-doping, since electron-hole exchange involving bare Coulomb matrix elements is the main source, see also the discussion of electron-hole exchange in the Appendix. Hence, environmental screening slows down the direct-to-indirect transition process only slightly.
%
\begin{figure}[h!t]
\centering
\includegraphics[width=.45 \textwidth]{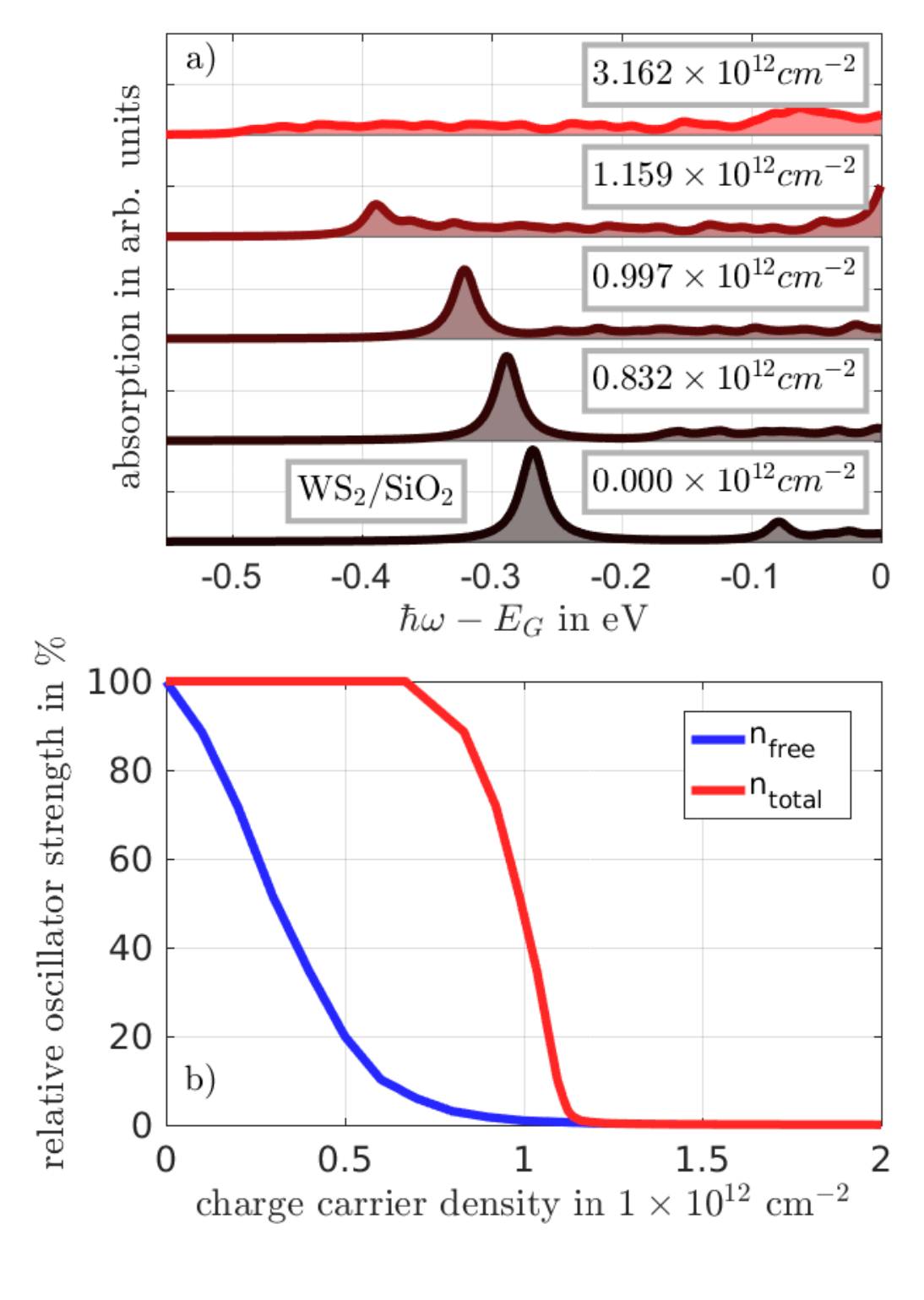}
\caption{\textbf{a)} Absorption spectra of \ws{} on \sio{}  at $T=300$ K given for total excited-carrier densities that include bound and quasi-free (unbound) electrons and holes. \textbf{b)} Relative oscillator strength of the A-exciton as a function of the total carrier density (red line) and the free carrier density (blue line). The free-carrier fraction of the total density is calculated assuming a quasi-equilibrium situation.}
\label{fig:abs_on_sio2}
\end{figure}
Like for the single-particle properties, renormalizations of the optical spectra in the presence of substrates are in general similar to the case of freestanding TMDC monolayers, as excitons still maintain a redshift under increasing carrier density.
\par
\begin{table}
\begin{ruledtabular}
\begin{tabular}{cccccccccc}
& \mos{} & \mose{} & \ws{} & \wse{} &\\ 
\hline
$E_B^A$ (meV) & 588 & 527 & 557 &508 \\
 
$E_B^B$ (meV)& 601 & 543 & 593 & 547 \\
&&&&\\
& \mos{}/\sio{} & \mose{}/\sio{} & \ws{}/\sio{} & \wse{}/\sio{}\\
\hline
$E_B^A$ (meV) & 301 & 288 & 268 & 253 \\
$E_B^B$ (meV) & 312 & 301 & 295 & 283 \\ 


\end{tabular}
\end{ruledtabular}
\label{tbl:results}
\caption{Results for exicton binding energies of the respective 1s-state for all four investigated materials with and without \sio{} substrate.}
\end{table}
In general, excited carriers can be present in the form of an unbound (quasi-free) electron-hole plasma or as bound excitons \cite{steinhoff_exciton_2017,steinleitner_direct_2017}. Above the Mott transition bound excitons cannot exist whereas below the Mott transition both phases can coexist. In this case, we in principle have to take into account renormalization effects due to excitons and unbound carriers separately. The contribution of bound excitons to single-particle renormalizations and screening is typically much smaller than contributions from unbound electrons and holes. Hence our results should represent a good approximation as long as all carrier densities are interpreted as the ionized fraction of a total electron-hole pair density that includes both phases. We illustrate this effect using the example of monolayer WS$_2$ on a SiO$_2$ substrate by rescaling the densities used to solve the SBE according to the fraction of ionized carriers given in Ref.~\citenum{steinhoff_exciton_2017}. Thereby we relate our results to the total electron-hole pair density that contains quasi-free carriers and carriers bound as excitons. The results shown in Fig.~\ref{fig:abs_on_sio2} exhibit a modified density-dependence due to the rescaling: Up to $1\times10^{12}$ cm$^{-2}$ only a weak line shift is visible, since most carriers are bound as excitons at intermediate densities, while strong redshifts appear between $1\times10^{12}$ cm$^{-2}$ and $3\times10^{12}$ cm$^{-2}$ before excitons dissociate at the Mott density. This behavior is in good agreement with recent experiments \cite{sie_observation_2017} except in the regime of high excitation, where a blueshift of the exciton resonance due to exciton-exciton interaction is observed. The rescaling of carrier density relies on the assumption of a quasi-equilibrium situation of photoexcited carriers, which is valid in a time window where excitons have formed out of excited electron-hole pairs and recombination has not yet set in. In this case, below the Mott transition, the fraction of unbound electrons and holes is determined essentially by exciton binding energies, the total electron-hole pair density, and the temperature \cite{steinleitner_direct_2017}. 
The consistent theoretical treatment of the coexisting exciton and plasma phases requires to go beyond the GW-approximation.

\section{Conclusion}

We have used the SBE approach to analyze single- and two-particle properties of \mos{}, \mose{}, \ws{} and \wse{} under optical excitation of electrons and holes as well as under doping of either carrier species.
\par
The many-body band-structure renormalizations under various excitation conditions are determined. Their magnitude is a combined effect of the orbital character of Bloch states and electron and hole populations in the different band-structure valleys. The resulting overall tendency is a direct-to-indirect transition of all considered monolayer TMDCs, which is mainly driven by the electron-hole exchange interaction. This implies a loss of charge carriers from the K- to the $\Sigma$-valley. The loss is more significant for the molybdenum-based compounds, where the energetically lowest interband transition at K is optically bright. Hence if the system is highly excited, the advantage of a semiconducting monolayer with a direct band gap slowly disappears. In particular, the $\Sigma$-valley has to be taken into account explicitely when describing the materials under the influence of excited carriers. Similar to strain effects, this tendency is expected to cause a quenching of PL intensity for high excitations\cite{steinhoff_efficient_2015} and to hamper population inversion at the K point.
\par
Due to the carrier-population dependence of electron-hole exchange, p- and n-doping lead to different renormalizations. For p-doping the direct-to-indirect transition is more pronounced as holes gather mostly in the K-valley. They drive a strong blue shift of the conduction-band states at K via electron-hole exchange, while there is no $\Sigma$-valley for holes and therefore no corresponding blue shift. The relative valley shifts at K and $\Sigma$ are comparably weak for n-doping as there is no hole population. These conduction-band renormalization effects have strong implications for the carrier densities at which a Lifshitz transition enabling phonon-driven superconductivity is expected. A comparison of the explicit carrier-carrier interaction results to DFT calculations reveals that for n-doping the relative valley shifts in DFT are overestimated at high densities, while they are underestimated at low densities. To correctly predict critical carrier densities for phonon-driven superconductivity in monolayer TMDCs, electron-hole exchange and/or plasma screening of electron-electron exchange has to be taken into account. Our findings suggest that in n-doped monolayer TMDCs slight photodoping would facilitate a Lifshitz transition via electron-hole exchange, which might be a means to optically induce superconductivity in two-dimensional materials. On the other hand, p-doping leads to an enhancement of the effective valence-band spin-orbit splitting at the K point by tens of meV due to different renormalizations of the upper and lower valence band.
\par
The competition between the shrinkage of the quasi-particle band gap and the plasma-screening-induced reduction of the exciton binding energy leads to a redshift of exciton resonances on the order of $100$ meV, while bleaching is observed due to phase space filling. The intricate interplay of many-particle renormalizations in the presence of excited carriers remains strong also in the presence of substrates and environmental screening effects. Our results are, therefore, highly relevant for any situation that involves doping and/or electrical or optical excitation of TMDC semiconductors.

\begin{acknowledgments}

We would like to thank Malte R\"osner for DFT band structure calculations and for fruitful discussions. This work has been supported by the Deutsche Forschungsgemeinschaft through the graduate school Quantum Mechanic Materials Modelling and through a grant for CPU time at the HLRN (Hannover/Berlin).

\end{acknowledgments}

\section{Appendix}

\subsection{Screening}

For excited semiconductors, we have to take into account two contributions to screening described by the polarization $\Pi=\Pi_{\textrm{b}}+\Pi_{\textrm{exc}}$, which acts as a self-energy for the screened Coulomb potential. The first term is the ``background'' polarization of the unexcited semiconductor while the second is due to electrons and holes that are introduced by doping or optical excitation. The contributions to screening are described by longitudinal dielectric functions $\varepsilon_{\textrm{b}}$ and $\varepsilon_{\textrm{exc}}$, respectively. By virtue of the background dielectric function, the bare Coulomb interaction $U$ is turned into the screened interaction $V$ according to the Dyson-type matrix equation \cite{rosner_wannier_2015, groenewald_valley_2016}
\begin{equation}
\begin{split} 
  V^{\alpha\beta}_{\bq}(\omega)&= \Big(U^{-1}_{\bq}-\Pi_{\textrm{b},\bq}(\omega)\Big)^{-1,\alpha\beta} \\
                               &= \sum_{\gamma}U^{\alpha\gamma}_{\bq}\Big(\mathbf{1}-U_{\bq}\Pi_{\textrm{b},\bq}(\omega)\Big)^{-1,\gamma\beta} \\
                               &=\sum_{\gamma}U^{\alpha\gamma}_{\bq}\varepsilon_{\textrm{b},\bq}^{-1,\gamma\beta}(\omega)
    \label{eq:V_matrix}
\end{split}
\end{equation}
with Coulomb matrix elements $U^{\alpha\beta}_{\bq}=U^{\alpha\beta\beta\alpha}_{\bq}$ calculated in a localized Wannier orbital basis $\ket{\bk,\alpha}$ where $\bk$ is a wave vector from the first Brillouin zone. The dependence of $U$ on only one quasi-momentum is due to the assumption that the Wannier functions are strictly localized to a single unit cell. The orbital basis may be chosen to represent the desired subspace of valence and conduction bands denoted by $\lambda$ via a unitary transformation \cite{steinhoff_influence_2014}
\begin{eqnarray}
U^{\lambda_1 \lambda_2 \lambda_3 \lambda_4}_{\mathbf{k}_1 \mathbf{k}_2 \mathbf{k}_3 \mathbf{k}_4} = \sum_{\alpha, \beta} \Big( c_{\alpha, \mathbf{k}_1}^{\lambda_1} \Big)^{*} \Big( c_{\beta, \mathbf{k}_2}^{\lambda_2} \Big)^{*} c_{\beta, \mathbf{k}_3}^{\lambda_3} c_{\alpha, \mathbf{k}_4}^{\lambda_4} U^{\alpha\beta\beta\alpha}_{|\mathbf{k}_1-\mathbf{k}_4|}\,,
\label{eq:v_abcd_appendix}
\end{eqnarray}
where $c_{\alpha, \mathbf{k}}^{\lambda}$ is the expansion coefficient of the Bloch state $\ket{\bk,\lambda}$ in the orbital basis. In the following we assume that Coulomb matrix elements are independent of carrier spin. The Coulomb interaction is further modified by excited-carrier screening thus yielding the fully screened interaction $W$:
\begin{equation}
\begin{split} 
  W^{\alpha\beta}_{\bq}(\omega)&= \Big(U^{-1}_{\bq}-\Pi_{\bq}(\omega)\Big)^{-1,\alpha\beta} \\
                               &= \Big(V^{-1}_{\bq}(\omega)-\Pi_{\textrm{exc},\bq}(\omega)\Big)^{-1,\alpha\beta} \\
                               &= \sum_{\gamma}V^{\alpha\gamma}_{\bq}(\omega)\Big(\mathbf{1}-V_{\bq}(\omega)\Pi_{\textrm{b},\bq}(\omega)\Big)^{-1,\gamma\beta} \\
  &=\sum_{\gamma}V^{\alpha\gamma}_{\bq}(\omega)\varepsilon^{-1,\gamma\beta}_{\textrm{exc},\bq}(\omega)\,.
    \label{eq:W_matrix}
\end{split}
\end{equation}
%
In a similar way, the electron self-energy can be split into background and excited-carrier contributions, $\Sigma=\Sigma_{\textrm{b}}+\Sigma_{\textrm{exc}}$, for which we use the GW scheme. The simplest approximation to the background polarization that is moreover consistent with this choice of self-energy \cite{baym_conservation_1961} is the Random Phase Approximation (RPA), in which we obtain \cite{groenewald_valley_2016}
\begin{equation} 
\begin{split}  
\Pi^{\alpha\beta}_{\textrm{b},\bq}(\omega)=&\frac{1}{\mathcal{A}}\sum_{\lambda,\lambda',\bk}\Big( c_{\alpha, \mathbf{k}}^{\lambda} \Big)^{*} \Big( c_{\beta, \mathbf{k}-\mathbf{q}}^{\lambda'} \Big)^{*} c_{\beta, \mathbf{k}}^{\lambda} c_{\alpha, \mathbf{k}-\mathbf{q}}^{\lambda'} \\
&\frac{f_{\bk-\bq}^{\lambda'}-f_{\bk}^{\lambda}}{\varepsilon_{\bk-\bq}^{\textrm{DFT},\lambda'}-\varepsilon_{\bk}^{\textrm{DFT},\lambda}+\hbar\omega+i\gamma}~.
 \label{eq:lindhard_b}
 \end{split}
\end{equation}
$\mathcal{A}$ is the crystal area, $f_{\bk}^{\lambda}$ are the electron occupancies taking the value $1$ for valence and $0$ for conduction bands, respectively, and $\varepsilon^{\textrm{DFT},\lambda}_{\bk}$ are single-particle energies without many-body renormalizations as approximately obtained from density functional theory (DFT) \cite{shishkin_implementation_2006}. As the background polarization includes only inter-band transitions, it is well-justified to work with the corresponding dielectric function in the static limit given by $\omega=0$ when describing intraband processes involving excited electrons and holes. Material-realistic static background dielectric functions for TMDC monolayers are obtained as described in Ref.~\citenum{steinhoff_exciton_2017}. The frequency dependence becomes important for energies above the single-particle band gap, which we discuss for a specific example in the next section. The dielectric function $\varepsilon_{\textrm{exc}}$ due to excited carriers is defined by Eq.~(\ref{eq:W_matrix}):
\begin{equation} 
\begin{split}  
\varepsilon_{\textrm{exc},\bq}^{\alpha\beta}(\omega) = \delta_{\alpha\beta}-\sum_{\gamma}V^{\alpha\gamma}_{\bq}(\omega)\Pi^{\gamma\beta}_{\textrm{exc},\bq}(\omega)~.
 \label{eq:eps_p_exc}
 \end{split}
\end{equation}
To simplify numerical calculations, we use a macroscopic dielectric function by setting $\varepsilon^{-1,\alpha\beta}_{\bq}(\omega)=\varepsilon^{-1}_{\bq}(\omega)\delta_{\alpha,\beta}$ and using a static Coulomb potential $V_{\bq}$ that is averaged over all elements of the Coulomb matrix in orbital representation. This corresponds to neglecting local-field effects in the dielectric function, which is justified for a plasma of excited electrons and holes that behave like quasi-free carriers\cite{hanke_local-field_1975}. In this case, we obtain the familiar Lindhard formula
%
\begin{equation} 
\begin{split}  
& \varepsilon_{\textrm{exc},\bq}(\omega) = \\ & 1-V_{\bq}\frac{1}{\mathcal{A}}\sum_{\lambda,\bk}
\frac{f_{\bk-\bq}^{0,\lambda}-f_{\bk}^{0,\lambda}}{\varepsilon_{\bk-\bq}^{0,\lambda}-\varepsilon_{\bk}^{0,\lambda}+\hbar\omega+i\gamma}\,.
 \label{eq:lindhard}
 \end{split}
\end{equation}
The energies $\varepsilon^{0,\lambda}_{\bk}$ contain only renormalization effects due to carriers in the ground state as described by the Green function $G^{0,\lambda}_{\bk}$. The latter is determined by the self-energy $\Sigma_{\textrm{b}}$ via a Dyson equation in quasi-particle approximation
\begin{equation} 
\begin{split}  
G^{0,\lambda,-1}_{\bk}(\omega)=G^{\textrm{DFT},\lambda,-1}_{\bk}(\omega)-\Sigma^{\lambda}_{\textrm{b},\bk}(\omega)\Big|_{\omega=\varepsilon_{\bk}^{0,\lambda}/\hbar}
 \label{eq:dyson_b}
 \end{split}
\end{equation}
%
and obtained for example from a first-principle GW calculation together with $V$. The corresponding electron and hole occupancies $f^{0,\lambda}_{\bk}$ are given by Fermi functions that are determined by the temperature as well as electron and hole chemical potentials in the quasi-equilibrium state under consideration. We neglect inter-band processes in Eq.~(\ref{eq:lindhard}) as the dielectric response due to excited electrons and holes is dominated by intra-band processes that lead for example to plasmonic phenomena. The static limit is obtained by setting $\omega=0$, and the dielectric function in Debye approximation follows the long-wavelength limit $\bq\rightarrow 0$:
\begin{equation} 
\begin{split}  
\varepsilon^{\textrm{ret}}_{\bq\rightarrow 0} = 1+V_{\bq}\sum_{\lambda}\sum_{\boldsymbol{\nu}}\frac{m^{\lambda}_{\boldsymbol{\nu}}f_{\boldsymbol{\nu}}^{\lambda}}{2\pi\hbar^2}
 \label{eq:debye}
 \end{split}
\end{equation}
with effective masses $m^{\lambda}_{\boldsymbol{\nu}}$ in the valley $\boldsymbol{\nu}$.

\subsection{Semiconductor Bloch equations in GW-approximation}
\begin{figure}[h!t]
\centering
\includegraphics[width=.35\textwidth]{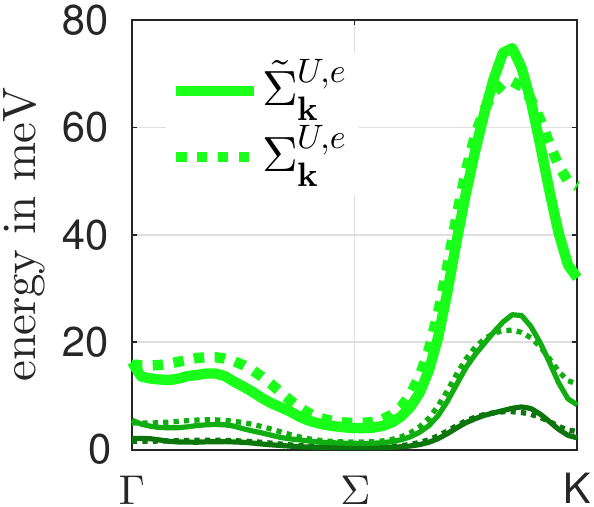}
\caption{Renormalization shifts of the conduction band due to hole population for freestanding MoS$_2$ along the $\Gamma$-K direction at $T=300$ K with (solid lines) and without (dashed lines) correlation contributions are shown for increasing charge-carrier densites $( 3\times10^{12}, 1\times10^{13},3\times10^{13}$ cm$^{-2} )$.}
\label{fig:screenedU}
\end{figure}

We decribe the two-particle properties of excited TMDC semiconductors using the semiconductor Bloch equations (SBE) for the microscopic interband polarizations $\psi_{\mathbf{k}}^{he}(t)=\left\langle a_{\mathbf{k}}^{h}\, a_{\mathbf{k}}^{e}\right\rangle(t)=-i\hbar G^{<,eh}_{\bk}(t)$ in the electron-hole picture. 
In this section we derive and discuss the SBE, as well as the corrsponding quasi-particle renormalizations, on the basis of a frequency-dependent GW self-energy. A static version of the BSE that is computationally less demanding is given in the subsequent section. The SBE can be derived using nonequilibrium Green function techniques \cite{jahnke_linear_1997, schafer_semiconductor_2002, manzke_quantum_2003}, assuming two conduction and two valence bands to be included in the excited-carrier self-energy $\Sigma_{\textrm{exc}}$:
\begin{widetext}
\begin{eqnarray}
\Big(i\hbar\frac{\mathrm{d}}{\mathrm{d}t} - \varepsilon^{\textrm{HF},h}_{\bk}(t)-\varepsilon^{\textrm{HF},e}_{\bk}(t)+i\gamma^{\textrm{El-Ph}}\Big) \psi_{\bk}^{he}(t) + \Big(1-f_{\bk}^{e}(t)-f_{\bk}^{h}(t)\Big) \Big(\mathbf{d}_{\bk}^{eh} \mathbf{E}(t)+\frac{1}{\mathcal{A}} \sum_{\bk'} V_{\bk\bk'\bk\bk'}^{ehhe} \psi_{\bk'}^{he}(t) \Big) \nonumber
\end{eqnarray}
\begin{eqnarray}
&= \sum\limits_{\lambda}^{} \int\limits_{-\infty}^{t} dt' \Big[ & \phantom{+} \, \Sigma^{>,e\lambda}_{\bk}(t,t')G^{<,\lambda h}_{\bk}(t',t)-\Sigma^{<,e\lambda}_{\bk}(t,t')G^{>,\lambda h}_{\bk}(t',t) \nonumber\\
&&+\,G^{<,e\lambda}_{\bk}(t,t')\Sigma^{>,\lambda h}_{\bk}(t',t)-G^{>,e\lambda}_{\bk}(t,t')\Sigma^{<,\lambda h}_{\bk}(t',t)\phantom{++} \Big]\,.
\label{eq:SBE_full}
\end{eqnarray}
\end{widetext}
The SBE are equivalent to a Dyson equation schematically given by $G^{-1}=G_0^{-1}+\mathbf{d}\cdot\mathbf{E}-\Sigma_{\textrm{exc}}$ for the Schwinger-Keldysh Green function $G$ that takes the place of the causal Green function in zero-temperature theory.
In Eq.~(\ref{eq:SBE_full}), $f_{\bk}^{\lambda}(t)=-i\hbar G^{<,\lambda}_{\bk}(t,t)$ are single-particle state occupancies and $\mathbf{d}_{\bk}^{eh}$ is the dipole matrix element describing light-matter interaction with the semiclassical electric field $E(t)$. $\varepsilon^{\textrm{HF},\lambda}_{\bk}(t)=\varepsilon_{\bk}^{0,\lambda}+\Sigma_{\bk}^{\textrm{H},\lambda}(t)+\Sigma_{\bk}^{\textrm{F},\lambda}(t)$ are single-particle band structures in Hartree-Fock approximation, while the collision terms on the right hand side account for many-body interaction effects beyond Hartree-Fock. The collision terms, which contain two-time Green functions $G^{\lessgtr}_{\bk}(t,t')$, are evaluated in GW-approximation to describe dephasing and higher-order renormalization contributions due to carrier-carrier interaction using the self-energy
\begin{equation}
\begin{split}
\Sigma^{\lessgtr,\lambda_1\lambda_2}_{\bk}(t,t')=i\hbar\frac{1}{\mathcal{A}}\sum_{\bk'\lambda_3\lambda_4}G^{\lessgtr,\lambda_3\lambda_4}_{\bk'}(t,t') W^{\gtrless,\lambda_1\lambda_4\lambda_2\lambda_3}_{\bk\bk'\bk\bk'}(t',t)\,.
\label{eq:GW}
\end{split}
\end{equation}
%
Dephasing due to carrier-phonon interaction is included via a phenomenological constant $\gamma^{\textrm{El-Ph}}$. The single-particle occupancies in general obey their own equations of motion. In the limit of weak external optical fields, however, they are assumed to be in a stationary quasi-equilibrium state with given temperature and carrier density that is either generated by prior optical excitation of the system or by doping with one carrier species. At the same time, in the weak-field limit, we may neglect all contributions in the collision terms that are nonlinear in interband polarizations. Then the SBE are equivalent to the well-known Bethe-Salpeter equation for excitons in dynamically screened ladder approximation \cite{bornath_two-particle_1999,kremp_quantum_2005}. Considering a quasi-equilibrium state, it is convenient to transform the SBE into frequency space. To simplify the equations, we use a quasi-particle approximation for the single-particle Green functions with self-consistently determined quasi-particle energies that are given by
\begin{equation}
\begin{split}
&\varepsilon^{\lambda}_{\bk} = \\ &\varepsilon_{\bk}^{0,\lambda}+\Sigma_{\bk}^{\textrm{H},\lambda}+\Sigma_{\bk}^{\textrm{F},\lambda}+\textrm{Re}\, \Sigma_{\bk}^{\textrm{MW},\textrm{ret},\lambda}(\omega)\Big|_{\omega=\varepsilon^{\lambda}_{\bk}/\hbar} \,.
\label{eq:quasip}
\end{split}
\end{equation}
The corresponding quasi-particle broadening follows from the imaginary part of the Montroll-Ward self-energy:
\begin{equation}
\begin{split}
&\Gamma^{\lambda}_{\bk} = -\textrm{Im}\, \Sigma_{\bk}^{\textrm{MW},\textrm{ret},\lambda}(\omega)\Big|_{\omega=\varepsilon^{\lambda}_{\bk}/\hbar} \,.
\label{eq:quasip_broadening}
\end{split}
\end{equation}
The Hartree and Fock self-energies, $\Sigma_{\bk}^{\textrm{H},\lambda}$ and $\Sigma_{\bk}^{\textrm{F},\lambda}$, describe the instantaneous interaction of the single-particle state $\ket{\bk,\lambda}$ with all other conduction- and valance-band states, where the exchange interaction is limited to states with equal spin. Due to their instantaneous nature, these self-energies are frequency-independent, as can be seen for the electron-electron exchange:
\begin{equation}
\begin{split} 
    \Sigma_{\bk}^{\textrm{F},e}(\omega)\Big|_{\textrm{El}}&=\int_{-\infty}^{\infty}d\tau e^{i\omega\tau}\Sigma_{\bk}^{\textrm{F},e}(\tau)\Big|_{\textrm{El}} \\
    &=i\hbar\frac{1}{\mathcal{A}}\sum_{\bk'}V^{eeee}_{\bk\bk'\bk\bk'}\int_{-\infty}^{\infty}d\tau e^{i\omega\tau} \delta(\tau) G^{<,e}_{\bk'}(\tau) \\
    &=i\hbar\frac{1}{\mathcal{A}}\sum_{\bk'}V^{eeee}_{\bk\bk'\bk\bk'}G^{<,e}_{\bk'}(\tau=0) \\
    &=i\hbar\frac{1}{\mathcal{A}}\sum_{\bk'}V^{eeee}_{\bk\bk'\bk\bk'}\int_{-\infty}^{\infty}\frac{d\omega}{2\pi}G^{<,e}_{\bk'}(\omega) \\
    &=i\hbar\frac{1}{\mathcal{A}}\sum_{\bk'}V^{eeee}_{\bk\bk'\bk\bk'}\int_{-\infty}^{\infty}\frac{d\omega}{2\pi}(-f^{e}(\omega)A_{\bk'}^{e}(\omega)) \\
    &=-\frac{1}{\mathcal{A}}\sum_{\bk'}V^{eeee}_{\bk\bk'\bk\bk'}f^{\textrm{KMS},e}_{\bk'}
    \label{eq:Fock_ee}
\end{split}
\end{equation}
In the fifth line we used the Kubo-Martin-Schwinger (KMS) relation for the lesser Green function \cite{kremp_quantum_2005} involving the single-particle spectral function $A_{\bk}^{\lambda}(\omega)=2i\textrm{Im}\,G^{\textrm{ret},\lambda}_{\bk}(\omega)$ and the Fermi function $f^{\lambda}(\omega)$. This leads to single-particle occupancies $f^{\textrm{KMS},\lambda}_{\bk}$ that are modified with respect to Fermi functions due to many-particle interaction effects. To be consistent with the quasi-particle ansatz (\ref{eq:quasip}) used to evaluate the collision terms, the Hartree-Fock self-energy should involve Lorentzian spectral functions. Yet as the long-range tails of the Lorentzian spectral functions impede the numerical evaluation, we replace them by Gaussian spectral functions with the same FWHM to capture the effect of quasi-particle broadening on single-particle occupancies at elevated densities. The Montroll-Ward self-energy $\Sigma_{\bk}^{\textrm{MW},\textrm{ret},\lambda}(\omega)$ contains all contributions to the GW self-energy beyond the Fock term\cite{kremp_quantum_2005,steinhoff_exciton_2017}, the intra-band term being given by
\begin{equation}
\begin{split} 
  & \Sigma_{\bk}^{\textrm{MW},\textrm{ret},\lambda}(\omega)\Big|_{\textrm{intra}} = i\hbar\int_{-\infty}^{\infty}\frac{d\omega'}{2\pi}\\
            \frac{1}{\mathcal{A}}\sum_{\bk'}&\frac{(1-f^{\lambda}(\omega-\omega')+n_{\textrm{B}}(\omega'))2i\,\textrm{Im}\,W^{\textrm{ret},\lambda\lambda\lambda\lambda}_{\bk\bk'\bk\bk'}(\omega') }{\hbar\omega-\varepsilon^{\lambda}_{\bk'}+i\Gamma^{\lambda}_{\bk'}-\hbar\omega'}
    \label{eq:MW}
\end{split}
\end{equation}
in the electron-hole picture. $n_{\textrm{B}}(\omega)$ is the Bose function and $2i\,\textrm{Im}\,W^{\textrm{ret}}(\omega)$ is the plasmon spectral function describing the excitation spectrum of the electron-hole plasma in terms of the retarded Coulomb interaction $W^{\textrm{ret}}$, which corresponds to the fully screened interaction introduced in Eq.~(\ref{eq:W_matrix}). All dielectric functions discussed in the following are retarded quantites as well. In Eq.~(\ref{eq:MW}) and the collision terms we discuss below, a frequency-dependent Fermi function remains under the integral hampering the numerical evaluation of these self-energy contributions. Hence we systematically replace the variable frequency argument by the real part of the quasi-particle energy that minimizes the denominator, which yields $f^{\lambda}(\omega-\omega')\approx f^{\lambda}_{\bk'}$ in Eq.~(\ref{eq:MW}).
\\
So far we explicitely considered only intra-band renormalizations of the quasi-particle energy. However, inter-band renormalization terms play an important role in the direct-to-indirect gap transition in TMDC semiconductors that we discuss in the main text. The inter-band contribution to the GW self-energy for conduction-band electrons due to valence-band electrons is given by
\begin{equation}
\begin{split} 
  & \tilde{\Sigma}_{\bk}^{\textrm{U,ret},c}(\omega) = \\ & -\frac{1}{\mathcal{A}}\sum_{\bk'}U^{cvcv}_{\bk\bk'\bk\bk'}f^{v}_{\bk'} +i\hbar\int_{-\infty}^{\infty}\frac{d\omega'}{2\pi}\\
            \frac{1}{\mathcal{A}}\sum_{\bk'}&\frac{(1-f^{v}_{\bk'}+n_{\textrm{B}}(\omega'))2i\,\textrm{Im}\,W^{\textrm{ret},cvcv}_{\bk\bk'\bk\bk'}(\omega') }{\hbar\omega-\varepsilon^{v}_{\bk'}+i\Gamma^{v}_{\bk'}-\hbar\omega'}\,,
    \label{eq:GW_cv}
\end{split}
\end{equation}
where we replace the KMS occupancies by Fermi functions in the exchange term for simplicity. Here, the self-energy is split into a bare exchange term and a correlation term that contains background as well as excited-carrier contributions, which is formally equivalent to typical first-principle implementations of the GW self-energy \cite{shishkin_implementation_2006}. Introducing hole occupancies $f^h_{\bk}=1-f^v_{\bk}$ and assuming that all contributions belonging to $f^h_{\bk}=0$ are already included in the ground-state band structure via Eq.~(\ref{eq:dyson_b}) we obtain for the inter-band renormalization due to excited carriers:
\begin{equation}
\begin{split} 
  & \tilde{\Sigma}_{\bk}^{\textrm{U,ret},e}(\varepsilon^{e}_{\bk}) = \frac{1}{\mathcal{A}}\sum_{\bk'}f^{h}_{\bk'}\\ & \Big[U^{eheh}_{\bk\bk'\bk\bk'} +i\hbar\int_{-\infty}^{\infty}\frac{d\omega'}{2\pi}\frac{2i\,\textrm{Im}\,W^{\textrm{ret},eheh}_{\bk\bk'\bk\bk'}(\omega') }{\varepsilon^{e}_{\bk}+\varepsilon^{h}_{\bk'}+i\Gamma^{h}_{\bk'}-\hbar\omega'}\Big]\,.
    \label{eq:GW_cv_2}
\end{split}
\end{equation}
In the electron-hole picture, the bare inter-band exchange term obtains a positive sign due to the opposite charge of electrons and holes. The correlation term is sensitive to the screened Coulomb matrix element at energies larger than the quasi-particle band gap and can thus not be described in a static limit. To quantify the effect of the inter-band correlation, we assume that intra-band contributions to screening are small at these frequencies and explicitely take into account only background screening. We numerically calculate the background dielectric matrix for MoS$_2$ using the RPA polarization (\ref{eq:lindhard_b}) and DFT band structures as introduced in Ref.~\citenum{steinhoff_influence_2014} with a phenomenological quasi-particle broadening of $10$ meV. Although the DFT results involve only a relatively small subspace around the direct band gap, we assume that these bands yield the dominant polarization contribution at the frequencies of interest. Using the screened Coulomb interaction we evaluate the conduction-band renormalization, Eq.~(\ref{eq:GW_cv_2}), with and without correlations as shown in Fig.~\ref{fig:screenedU}. Without correlations the results correspond to those in the top panel of Fig.~\ref{fig:sx_ch} in the main text.
We find that while the correlations yield a correction of about 30 \% at the K-point, the overall behaviour of inter-band renormalizations is described already by the bare electron-hole exchange term. Hence for simplicity we approximate the inter-band renormalizations by
\begin{equation}
\begin{split} 
  & \Sigma_{\bk}^{\textrm{U},e} = \frac{1}{\mathcal{A}}\sum_{\bk'}U^{eheh}_{\bk\bk'\bk\bk'}f^{h}_{\bk'}
    \label{eq:GW_cv_2}
\end{split}
\end{equation}
and a corresponding term for valence-band energies which can both be included in the Hartree-Fock energies $\varepsilon^{\textrm{HF},\lambda}_{\bk}$.





Altogether, the SBE in frequency space are given by
\begin{equation}
\begin{split}
 \Big(\hbar\omega-\varepsilon^{\textrm{HF},h}_{\bk}-&\varepsilon^{\textrm{HF},e}_{\bk} -\Delta^{eh}_{\bk}(\omega) +i\gamma^{\textrm{El-Ph}}\Big)\psi_{\bk}^{he}(\omega) \\
 +\Big(1-f_{\bk}^{\textrm{KMS},e}-&f_{\bk}^{\textrm{KMS},h}\Big)\Big(\mathbf{d}_{\bk}^{eh}\cdot \mathbf{E}(\omega)+\frac{1}{\mathcal{A}}\sum_{\bk'}V_{\bk\bk'\bk\bk'}^{ehhe}\psi_{\bk'}^{he}(\omega)\Big) \\
 &+\sum_{\bk'}V_{\bk\bk'}^{\textrm{eff},eh}(\omega)\psi_{\bk'}^{he}(\omega) =0\,.
\label{eq:SBE_GW_freq}
\end{split}
\end{equation}
The excitation-induced correlations that cause spectral shifts and lifetime broadening of optical transitions are here described in GW-approximation by the frequency-dependent terms
\begin{equation}
\begin{split}
\Delta^{eh}_{\bk}(\omega)&=\Sigma_{\bk}^{\textrm{MW},\textrm{ret},e}(\hbar\omega-\varepsilon^h_{\bk}+i\Gamma^h_{\bk})\\&+\Sigma_{\bk}^{\textrm{MW},\textrm{ret},h}(\hbar\omega-\varepsilon^e_{\bk}+i\Gamma^e_{\bk})
\label{eq:Delta}
\end{split}
\end{equation}
and
\begin{equation}
\begin{split}
V_{\bk\bk'}^{\textrm{eff},eh}(\omega)&=i\hbar\int_{-\infty}^{\infty}\frac{d\omega'}{2\pi}\Bigg\{\\
    &\frac{(1-f^{h}_{\bk}+n_{\textrm{B}}(\omega'))2i V^{ehhe}_{\bk\bk'\bk\bk'}\textrm{Im}\,\varepsilon^{-1}_{\textrm{exc},\bk-\bk'}(\omega')}{\hbar\omega-\varepsilon^{h}_{\bk}-\varepsilon^{e}_{\bk'}+i\Gamma^h_{\bk}+i\Gamma^e_{\bk'}-\hbar\omega'} \\
   + &\frac{(1-f^{e}_{\bk}+n_{\textrm{B}}(\omega'))2i V^{ehhe}_{\bk\bk'\bk\bk'}\textrm{Im}\,\varepsilon^{-1}_{\textrm{exc},\bk-\bk'}(\omega')}{\hbar\omega-\varepsilon^{e}_{\bk}-\varepsilon^{h}_{\bk'}+i\Gamma^e_{\bk}+i\Gamma^h_{\bk'}-\hbar\omega'}\Bigg\}\,.
\label{eq:Veff}
\end{split}
\end{equation}
%

\subsection{Static limit of SBE}

As the SBE with full frequency dependence are computationally very demanding, we resort to a static approximation for most of the results shown in this paper. The validity of the static approximation is confirmed by comparing relative shifts of K- and $\Sigma$-valleys obtained from both theories for a representative case. In the static approximation, the frequency-dependence of dephasing is neglected, which simplifies the numerical evaluation considerably. This limit can be systematically derived by assuming that any excitation energy involving pairs of free (quasi-)particles, $\hbar\omega-\varepsilon^{\lambda}_{\bk}-\varepsilon^{\lambda'}_{\bk'}$, in the correlation integrals (\ref{eq:Delta}) and (\ref{eq:Veff}) is small compared to characteristic energies $\hbar\omega'$ occurring in the dielectric function. \cite{bornath_two-particle_1999, kremp_quantum_2005} Then we obtain for example
\begin{equation}
\begin{split}
\Delta^{eh}_{\bk}(\omega)&\approx -i\hbar\int_{-\infty}^{\infty}\frac{d\omega'}{2\pi} \frac{1}{\mathcal{A}}\sum_{\lambda \bk'} \\
    &\frac{(1-f^{\lambda}_{\bk'}+n_{\textrm{B}}(\omega'))2i V^{\lambda\lambda\lambda\lambda}_{\bk\bk'\bk\bk'}\textrm{Im}\,\varepsilon^{-1}_{\textrm{exc},\bk-\bk'}(\omega')}{\hbar\omega'}
    \,.
\label{eq:Delta_approx1}
\end{split}
\end{equation}
We use the relation 
\begin{equation}
\begin{split}
\mathcal{P} \int_{-\infty}^{\infty}\frac{d\omega'}{\pi}\frac{\textrm{Im}\,\varepsilon^{-1}_{\bq}(\omega')}{\omega'-\omega}=\textrm{Re}\,\varepsilon^{-1}_{\bq}(\omega)-1
\label{eq:KK}
\end{split}
\end{equation}
with the Cauchy principal value $\mathcal{P}$, corresponding to the dispersion relation for the electronic susceptibility, for $\omega=0$. Since $\textrm{Im}\,\varepsilon^{-1}_{\bq}(\omega)$ is an odd function of $\omega$, the integrand has no pole at $\omega'=0$ and the principal value becomes a regular integral. Furthermore, we use the fact that $n_{\textrm{B}}(\omega)+\frac{1}{2}$ is an odd function of $\omega$ as well, to obtain
\begin{equation}
\begin{split}
\int_{-\infty}^{\infty}\frac{d\omega'}{\pi}\frac{\textrm{Im}\,\varepsilon^{-1}_{\bq}(\omega')(n_{\textrm{B}}(\omega')+\frac{1}{2})}{\omega'}=0\,.
\label{eq:integral2}
\end{split}
\end{equation}
Combining Eq.~(\ref{eq:Delta_approx1}) with (\ref{eq:KK}) and (\ref{eq:integral2}), the intra-band correlation becomes
\begin{equation}
\begin{split}
\Delta^{eh}_{\bk}(\omega)&\approx \frac{1}{\mathcal{A}}\sum_{\lambda \bk'} \big( \frac{1}{2}-f_{\bk'}^{\lambda}\big)\left[W^{\lambda\lambda\lambda\lambda}_{\bk\bk'\bk\bk'}(\omega=0)- V^{\lambda\lambda\lambda\lambda}_{\bk\bk'\bk\bk'}\right]
\label{eq:Delta_approx2}
\end{split}
\end{equation}
with $W^{\lambda\lambda\lambda\lambda}_{\bk\bk'\bk\bk'}(\omega=0)=V^{\lambda\lambda\lambda\lambda}_{\bk\bk'\bk\bk'} \varepsilon^{-1}_{\textrm{exc},\bk-\bk'}(\omega=0) $. Similarly, the inter-band correlation is given by
\begin{equation}
\begin{split}
V_{\bk\bk'}^{\textrm{eff},eh}(\omega)&\approx \big(1-f_{\bk}^{h}-f_{\bk}^{e}\big)\left[W^{ehhe}_{\bk\bk'\bk\bk'}(\omega=0)- V^{ehhe}_{\bk\bk'\bk\bk'}\right]\,.
\label{eq:Veff_approx2}
\end{split}
\end{equation}
All correlation terms become frequency-independent and real, hence collision broadening of exciton lines is not contained in static approximation. We summarize collision broadening due to carrier-phonon and carrier-carrier interaction into the phenomenological constant $\gamma$. In conclusion, the SBE in static approximation are given by
\begin{equation}
\begin{split}
 \Big(\hbar\omega-\tilde{\varepsilon}^{h}_{\bk}-&\tilde{\varepsilon}^{e}_{\bk} +i\gamma\Big)\psi_{\bk}^{he}(\omega) \\
 +\Big(1-f_{\bk}^{e}-f_{\bk}^{h}\Big)&\Big(\mathbf{d}_{\bk}^{eh}\cdot \mathbf{E}(\omega)+\frac{1}{\mathcal{A}}\sum_{\bk'}W_{\bk\bk'\bk\bk'}^{ehhe}\psi_{\bk'}^{he}(\omega)\Big) =0\,.
\label{eq:SBE_SXCH}
\end{split}
\end{equation}
The renormalized energies $\tilde{\varepsilon}^{\lambda}_{\bk}$ are composed of ground-state band structures $\varepsilon^{0,\lambda}_{\bk}$, Hartree-Fock contributions including bare inter-band exchange and the intra-band correlation $\Delta^{eh}_{\bk}(\omega)$ in static limit:
\begin{equation}
\begin{split}
\tilde{\varepsilon}^{\lambda}_{\bk}&=\varepsilon^{0,\lambda}_{\bk}+\Sigma_{\bk}^{\textrm{H},\lambda}+\Sigma_{\bk}^{\textrm{U},\lambda}+\Sigma_{\bk}^{\textrm{SX},\lambda}+\Sigma_{\bk}^{\textrm{CH},\lambda} \\
&=\varepsilon^{0,\lambda}_{\bk}+\Sigma_{\bk}^{\textrm{H},\lambda}+\frac{1}{\mathcal{A}}\sum_{\bk'\lambda'\neq\lambda}U_{\bk\bk'\bk\bk'}^{\lambda\lambda'\lambda\lambda'}f^{\lambda'}_{\bk'}\\ 
&-\frac{1}{\mathcal{A}}\sum_{\bk'}W_{\bk\bk'\bk\bk'}^{\lambda\lambda\lambda\lambda}f^{\lambda}_{\bk'} 
+\frac{1}{2\mathcal{A}}\sum_{\bk'}\left[W^{\lambda\lambda\lambda\lambda}_{\bk\bk'\bk\bk'}- V^{\lambda\lambda\lambda\lambda}_{\bk\bk'\bk\bk'}\right]\,.
\label{eq:en_SXCH}
\end{split}
\end{equation}
The energies contain the screened-exchange interaction (SX) term and the Coulomb-hole term (CH) involving the difference of Coulomb potential screened and unscreened by excited carriers. Due to the last two terms, this approximation is also termed \textit{screened-exchange-Coulomb-hole} (SXCH) approximation.

\bibliographystyle{apsrev}


\end{document}